\documentclass[twocolumn]{aastex63}
\usepackage{CJKutf8}
\usepackage{hyperref}
\newcommand{\cntext}[1]{\begin{CJK}{UTF8}{gbsn}#1\end{CJK}\kern-1ex}
\newcommand{\uat}[2]{\href{http://astrothesaurus.org/uat/#2}{#1 (#2)}}
\shorttitle{Wei et al. 2021}
\shortauthors{Wei et al.}
\graphicspath{{./}{figures/}}
\turnoffeditone

\begin{document}

\title{Coronal Magnetic Field Measurements along a Partially Erupting Filament in a Solar Flare}

\author[0000-0002-6628-6211]{Yuqian Wei}
\affiliation{Center for Solar-Terrestrial Research, New Jersey Institute of Technology, 323 M L King Jr. Blvd., Newark, NJ 07102-1982, USA}

\author[0000-0002-0660-3350]{Bin Chen (\cntext{陈彬})}
\affiliation{Center for Solar-Terrestrial Research, New Jersey Institute of Technology, 323 M L King Jr. Blvd., Newark, NJ 07102-1982, USA}

\author[0000-0003-2872-2614]{Sijie Yu (\cntext{余思捷})}
\affiliation{Center for Solar-Terrestrial Research, New Jersey Institute of Technology, 323 M L King Jr. Blvd., Newark, NJ 07102-1982, USA}

\author{Haimin Wang}
\affiliation{Center for Solar-Terrestrial Research, New Jersey Institute of Technology, 323 M L King Jr. Blvd., Newark, NJ 07102-1982, USA}

\author{Ju Jing}
\affiliation{Center for Solar-Terrestrial Research, New Jersey Institute of Technology, 323 M L King Jr. Blvd., Newark, NJ 07102-1982, USA}

\author[0000-0003-2520-8396]{Dale E. Gary}
\affiliation{Center for Solar-Terrestrial Research, New Jersey Institute of Technology, 323 M L King Jr. Blvd., Newark, NJ 07102-1982, USA}

\correspondingauthor{Bin Chen, Haimin Wang}\email{bin.chen@njit.edu, haimin.wang@njit.edu}

\begin{abstract}

Magnetic flux ropes are the centerpiece of solar eruptions. Direct measurements for the magnetic field of flux ropes are crucial for understanding the triggering and energy release processes, yet they remain heretofore elusive. Here we report microwave imaging spectroscopy observations of an M1.4-class solar flare that occurred on 2017 September 6, using data obtained by the Expanded Owens Valley Solar Array. This flare event is associated with a partial eruption of a twisted filament observed in H$\alpha$ by the Goode Solar Telescope at the Big Bear Solar Observatory. The extreme ultraviolet (EUV) and X-ray signatures of the event are generally consistent with the standard scenario of eruptive flares, with the presence of double flare ribbons connected by a bright flare arcade.
Intriguingly, this partial eruption event features a microwave counterpart, whose spatial and temporal evolution closely follow the filament seen in H$\alpha$ and EUV. The spectral properties of the microwave source are consistent with nonthermal gyrosynchrotron radiation. Using spatially resolved microwave spectral analysis, we derive the magnetic field strength along the filament spine, which ranges from 600--1400 Gauss from its apex to the legs. The results agree well with the non-linear force-free magnetic model extrapolated from the pre-flare photospheric magnetogram. We conclude that the microwave counterpart of the erupting filament is likely due to flare-accelerated electrons injected into the filament-hosting magnetic flux rope cavity following the newly reconnected magnetic field lines.   

\end{abstract}

\keywords{
\uat{Solar magnetic fields}{1503}, \uat{Solar filaments}{1495}, \uat{Solar corona}{1483}, \uat{Solar flares}{1496}, \uat{Non-thermal radiation sources}{1119}, \uat{Solar radio emission}{1522}, \uat{Solar radio flares}{1342}}

\section{Introduction} \label{sec:intro}
Magnetic flux ropes (MFRs) are the key to understanding solar eruptions \citep{Filippov2015}. Since the magnetic field plays a dominant role in the low-plasma-$\beta$ environment in the low solar corona, measurements of the magnetic properties of MFRs are crucial for understanding their triggering and the associated energy release processes, leading to major solar activities \citep{Liu2016}. 
 
To date, the most commonly used method to infer the magnetic field of MFRs is through nonlinear force-free field (NLFFF) extrapolations \citep{Wiegelmann2008}. While the NLFFF method has provided important insights into the magnetic topology and, in some cases, the evolution of the MFRs \citep[e.g.,][]{Kliem2013, Inoue2016, Guo2019}, it is, after all, an indirect method with intrinsic limitations \citep[see., e.g.,][for discussions]{Metcalf2008, 2009ApJ...696.1780D}.
Direct measurements of the coronal magnetic field based on the Zeeman effect, Hanle effect, or a combination of both, have been performed by using polarization measurements of optical or infrared (IR) lines \citep{Lin_2004, Gibson2016, Raouafi2016}. Occasionally, this technique has been applied to the measurements of the magnetic field of prominences/filaments \citep{Bommier1994,Merenda2006} and coronal rain \citep{2019ApJ...874..126K}. 
Linear polarization of IR forbidden lines are also used to probe the magnetic structure of coronal cavities \citep{Dove2011, Bak-Steslicka2013}. \edit1{Recently, there has also been success in constraining the coronal magnetic field by using certain EUV lines sensitive to the magnetically induced transition (MIT) effect \citep{2021ApJ...915L..24B, 2021ApJ...918L..13C,2021ApJ...913....1L}.} However, these measurements are often limited by the signal-to-noise ratio and require a \edit1{relatively} long integration time. For instance, with current optical/IR instrumentation, it typically takes tens of minutes for linear polarization and hours for circular polarization \citep{Bak-Steslicka2013}. Although the required integration time will be reduced by an order of magnitude with the operation of the Cryo-NIRSP spectropolarimeter at the Daniel K. Inouye Solar Telescope \citep{Harrington2017} in the near future (thanks to its large collecting area), it will remain difficult to track the rapidly evolving magnetic field of MFRs at a time scale of order 1--10 s in solar eruptions and flares. In addition, because the spectral lines from the coronal plasma are orders of magnitude weaker than their counterpart from the photosphere, such measurements can only be made above the solar limb. 

\edit1{Waves and oscillations have also been used to diagnose the coronal magnetic field \citep[see, e.g., reviews by][and references therein]{Nakariakov2005, Andries2009}. Despite recent success to obtain spatially resolved maps of the coronal magnetic field above sunspots \citep{Jess2016} and of the quiescent corona \citep{Yang2020}, it remains difficult to diagnose the rapidly evolving and complex magnetic structures in the flaring region.}

Microwave spectral diagnostics provide another means for measuring the coronal magnetic field both over the limb and against the disk. The microwave emission arises from thermal or nonthermal electrons gyrating in the magnetic field, producing gyroresonance or gyrosynchrotron radiation with spectral properties sensitive to the magnetic field strength and direction \citep{Gary2004}. Thanks to the operation of the Expanded Owens Valley Solar Array (EOVSA) \citep{Nita2016, Gary2018}, significant progress has been made in measuring the dynamically evolving coronal magnetic field in solar flares using spatially resolved microwave spectroscopy at a high, 1-s cadence. The technique of faithfully reconstructing a coronal magnetic field map of a flare arcade has been previously demonstrated by \citet{Gary2013} using a three-dimensional model arcade filled with nonthermal electrons. With EOVSA data, a fast decay of the coronal magnetic field in the cusp region above the flare arcade was first reported by \citet{Fleishman2020}. In \citet{Chen2020a}, EOVSA imaging spectroscopy is used to derive the magnetic field profile along a large-scale reconnection current sheet trailing an erupting MFR, which matches very well with results from numerical simulations. Similar techniques can be used to derive spatially resolved measurements of the magnetic field along an erupting MFR. However, such measurements have not been realized heretofore due to the lack of suitable observational data. This work will fill that gap.

Although nonthermal counterparts of the erupting MFRs or coronal mass ejections (CMEs) have been occasionally reported in both radio \citep{Stewart1982,Gary1985,2001ApJ...558L..65B,Vrsnak2003a,2007ApJ...660..874M,2013ApJ...766..130T,2014ApJ...782...43B,2017A&A...608A.137C,2019A&A...623A..63M,Chen2020b,2020ApJ...893...28M,2021ApJ...906..132C} and X-ray wavelengths \citep{1992ApJ...390..687K,2001ApJ...561L.211H,2007ApJ...669L..49K,2013ApJ...779L..29G}, where the nonthermal electrons are accelerated and how they gain access to the MFR/CME cavity remain outstanding questions. In the low corona, it is often assumed that the electrons, presumably accelerated at or around the flare reconnection site, are injected into the MFR cavity following the newly reconnected field lines. In the upper corona, additionally, the CME-driven shocks may play an increasingly important role in particle acceleration. As such, transport of the shock-accelerated electrons (and ions) from downstream of the shock to the CME cavity would be important for understanding the associated nonthermal emissions. To elucidate these processes, detailed spectral imaging of these radio/X-ray sources with sufficient temporal and angular resolution would be particularly helpful. In a recent study by \citet{Chen2020b}, aided by spectral imaging enabled by EOVSA, microwave counterparts that outline the central region and the two conjugate footpoints of an erupting MFR cavity have been identified in the low corona ($<\!0.2 R_{\odot}$ above the surface) with an edge-on viewing perspective. Their similar light curves and spectral properties suggest that the three sources are likely associated with the same nonthermal electron population injected from the underlying reconnecting current sheet during the erupting of the MFR.

Here we report a multi-wavelength study of a GOES M1.4-classs solar flare on 2017 September 6, which is associated with a partial eruption of a filament. Different from the event in \citet{Chen2020b}, which has an edge-on viewing perspective over the limb, this erupting filament is viewed against the solar disk, thus giving us a unique opportunity to derive the coronal magnetic field \textit{along} the filament axis using both microwave spectral analysis and non-linear force-free field (NLFFF) extrapolations. 
After an overview of the multi-wavelength observations and a discussion of the flare context in Section~\ref{sec:multi}, we present the microwave imaging spectroscopy observations of the filament eruption in Section~\ref{sec:MW}. Also discussed there are the magnetic field measurements returned from the spatially resolved microwave spectral analysis in comparison to the NLFFF extrapolation results. In Section~\ref{sec:discussion}, we discuss the nature of the filament/MFR system revealed by the multi-wavelength observations, as well as a possible interpretation of the partial eruption.

\begin{figure*}[!ht]
\centering
\includegraphics[width=0.9\textwidth]{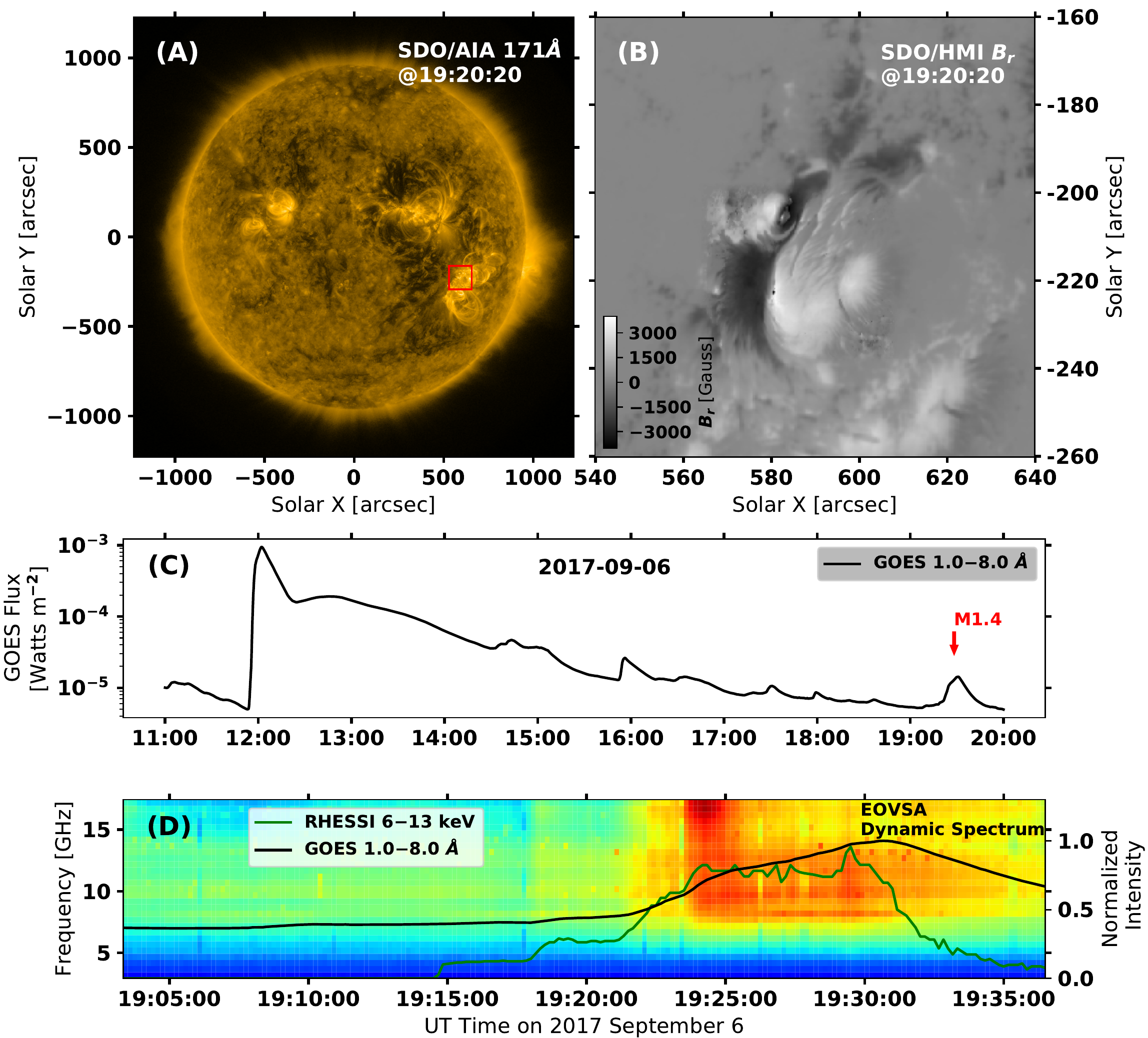}
\caption{\label{fig:over_all} (A) AR 12673 as observed in EUV by the SDO/AIA 171~\AA\ filterband on 2017 September 6 at 19:20:20 UT. (B) Detailed view of the SDO/HMI \edit1{radial field} magnetogram of the core region of the AR (red box in (A)). 
(C) GOES 1--8 \AA\ soft X-ray (SXR) light curve from 11 UT to 20 UT on 2017 September 6. The M1.4 flare event under study (marked by the red arrow) occurs during the late decay phase of the large X9.3 flare. (D) Background-subtracted EOVSA microwave dynamic spectrum from 19:03 UT to 19:37 UT. The black and green curves are for RHESSI 6--13 keV X-ray and GOES 1--8~\AA\ light curves, respectively.}
\end{figure*}

\begin{figure*}[!ht]
\centering
\includegraphics[width=0.9\textwidth]{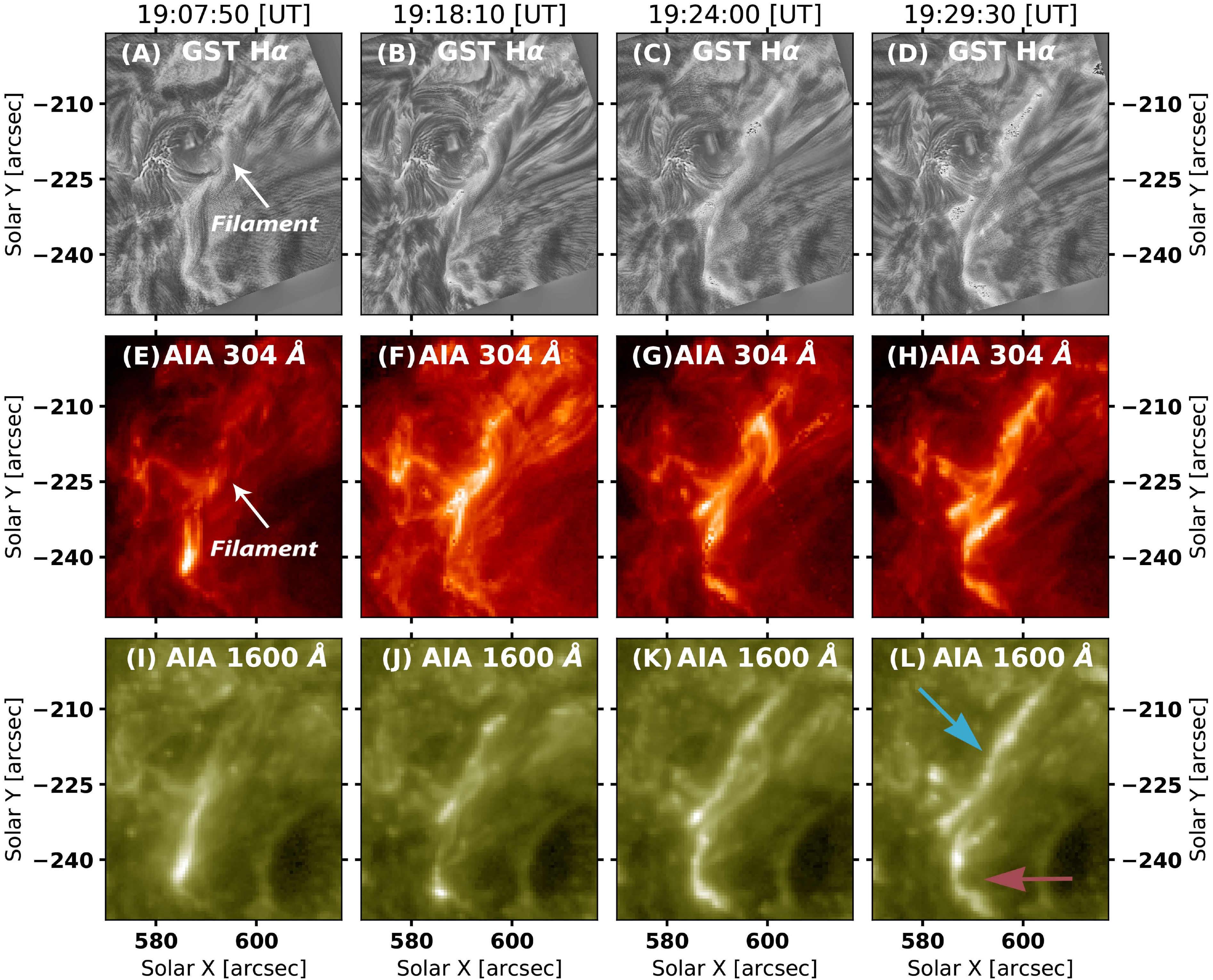}
\caption{\label{fig:filament} (A)--(D) BBSO/GST H$\alpha$ line center images at four selected times during the event. (E)--(H) SDO/AIA 304~\AA\ images. The dark filament, seen in both H$\alpha$ and 304~\AA, is marked by the white arrows. (I)--(L) SDO/AIA 1600~\AA\ images showing the development of the two flare ribbons (arrows in (L)). An animation of the full time evolution is included in Appendix as Figure~\ref{video:sdo_gst}.}
\end{figure*}

\section{Event Overview}\label{sec:multi}
\subsection{Multi-wavelength Data}
The flare under study took place in active region (AR) 12673, which had a record-breaking fast magnetic flux emergence \citep{Sun2017} and high photospheric and coronal magnetic field strength \citep{2019AAS...23440205W, Anfinogentov2019}. The event occurred about 7 hours after the peak of the X9.3 class flare (SOL2017-09-06T11:53, the largest flare in Solar Cycle 24) in the same AR. 
During the decay phase of the X9.3 flare, 4 M-class flares occur, including the SOL2017-09-06T19:29:30 M1.4 flare\footnote{The flare classes quoted here follow those reported by NOAA based on GOES 15 data. We note that the GOES 16 shown in Figure~\ref{fig:over_all}(C) reports a 1--8 \AA\ flux of $2.2\times 10^{-5}$ W m$^{-2}$ at the peak of this event, which would be named an M2.2 class. Such an inconsistency is due to a scaling factor applied to data prior to GOES 16, which is well documented by NOAA/NCEI (\url{https://www.ngdc.noaa.gov/stp/satellite/goes-r.html}).} under study here (Figure~\ref{fig:over_all}(C)). At the time, as shown in Figure~\ref{fig:over_all}(B), the photospheric magnetic field configuration of the AR appears as a quadrupolar configuration. 

The observation of the flaring region obtained by the 1.6-m Goode Solar Telescope of the Big Bear Solar Observatory (BBSO/GST; \citealt{Cao2010}) is available from 19:00 UT--20:09 UT. The Visible Imaging Spectrometer (VIS) at GST provides observation in H$\alpha$ 6563~\AA\ line center and line wing ($\pm$0.4~\AA\ and $\pm$0.8~\AA), with an angular resolution of $0''.1$ and a field-of-view (FOV) of $57''\times 64''$. The Helioseismic and Magnetic Imager (HMI; \citealt{2012SoPh..275..207S}) and the Atmospheric Imaging Assembly (AIA; \citealt{lemen2012}) on-board the Solar Dynamics Observatory (SDO; \citealt{pesnell2012}) provide full-disk magnetograms and multi-band extreme-ultraviolet (EUV) and ultraviolet (UV) images, respectively, with an angular resolution of $1''$--$1''.5$. The images from BBSO/GST are enhanced with the multi-scale Gaussian normalization (MGN) method \citep{Morgan2014}. The Reuven Ramaty High Energy Solar Spectroscopic Imager (RHESSI; \citealt{lin2002}) observed the event from 19:15 UT. The X-ray response from this flare can be detected against the background up to $\sim$13 keV. X-ray imaging reconstruction is performed during the flare peak using the standard CLEAN method \citep{Hurford2002} at 6--13 keV with an integration time of 120 s.

The M1.4 event is fully covered by EOVSA with 134 frequencies in 2.5--18 GHz over 31 evenly spaced spectral windows (referred to as SPW 0 to SPW 30). Phase calibration was done against a celestial source 1229+020. A self-calibration procedure in both phase and amplitude is performed based on a 4-s-averaged data around the flare peak. Most of the enhanced microwave emission associated with the flare is in the 5--18 GHz range, which will be the focus of our analysis in Section~\ref{sec:MW}. 

\begin{figure*}[!ht]
\centering
\includegraphics[width=0.9\textwidth]{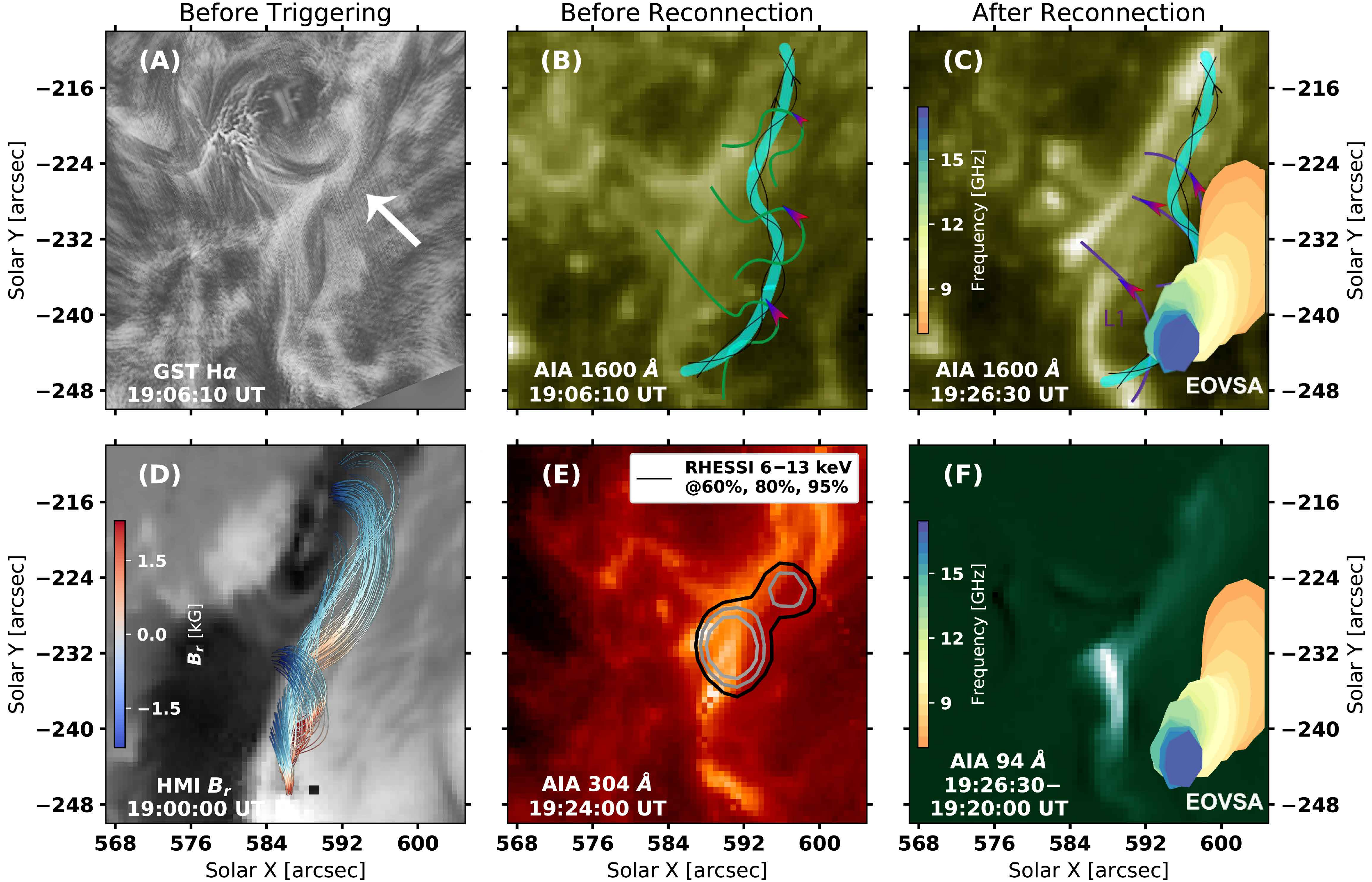}
\caption{\label{fig:story} 
(A) Filament as seen by BBSO/GST H$\alpha$ before the onset of the event at 19:06:10 UT (white arrow). (B) and (C) Schematic of the pre- and post-reconnection magnetic field lines (green and purple curves) induced by the rising filament. The background is the corresponding SDO/AIA 1600~\AA\ image showing the formation of the two bright flare ribbons. EOVSA 6.4--15.9 GHz images at 19:26:30 (pre-flare background subtracted) are also shown in (C) as color contours (90\% of the maximum). 
(D) Selected field lines near the PIL region derived from the NLFFF results based on the SDO/HMI vector magnetogram at 19:00 UT.
(E) RHESSI 6--13 keV X-ray source (60\%, 80\%, and 95\% of the maximum) overlaid on SDO/AIA 304~\AA\ EUV image during the impulsive phase. X-ray spectral analysis suggests that the source is associated with thermal bremsstrahlung emission from $\sim$28 MK plasma. (F) Base difference SDO/AIA 94~\AA\ image (19:26:30$-$19:20:00 UT) showing the bright post-reconnection flare arcade. EOVSA 6.4--15.9 GHz contours at 19:26:30 are also shown.}
\end{figure*}

\begin{figure*}[!ht]
\centering
\includegraphics[width=0.9\textwidth]{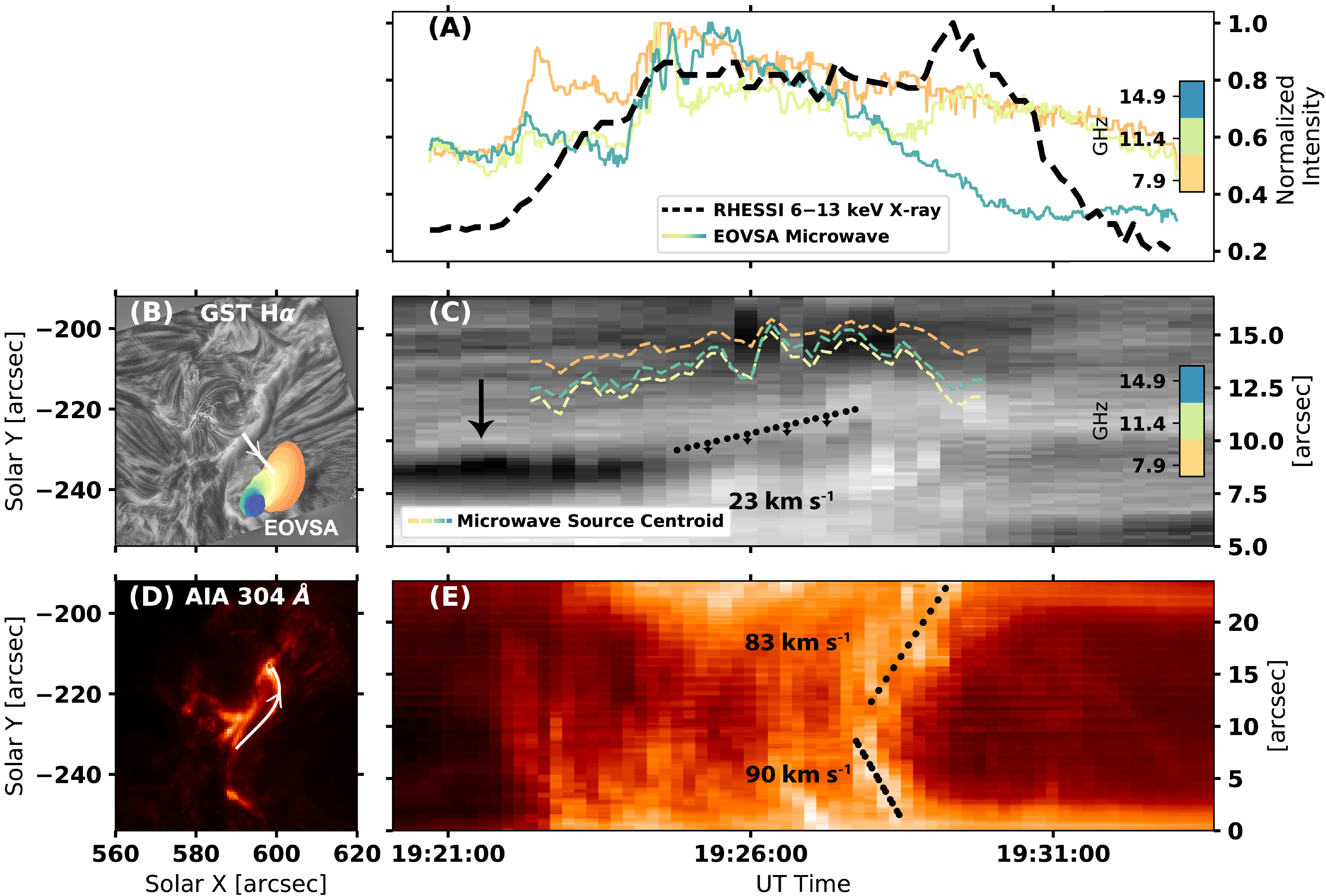}
\caption{\label{fig:stack} (A) X-ray and microwave light curves during the partial eruption of the filament.
(B) Reference BBSO/GST H$\alpha$ image at 19:20:20 UT overlaid with EOVSA 7.9--15.9 GHz contours (90\% of the maximum). (C) Time--distance stack plot of the BBSO/GST H$\alpha$ image series made along a slice as indicated by the white curve in (B). The color dashed curves indicate the location of EOVSA microwave centroid at three selected frequencies. The dark, rising filament is indicated by the black arrow. (D) Reference SDO/AIA 304~\AA\ image at 19:20:20 UT. (E) Time--distance stack plot made along the filament (white curve in (D)), showing the draining filament material following the partial eruption.}
\end{figure*}

\begin{figure*}
\centering
\includegraphics[width=0.95\textwidth]{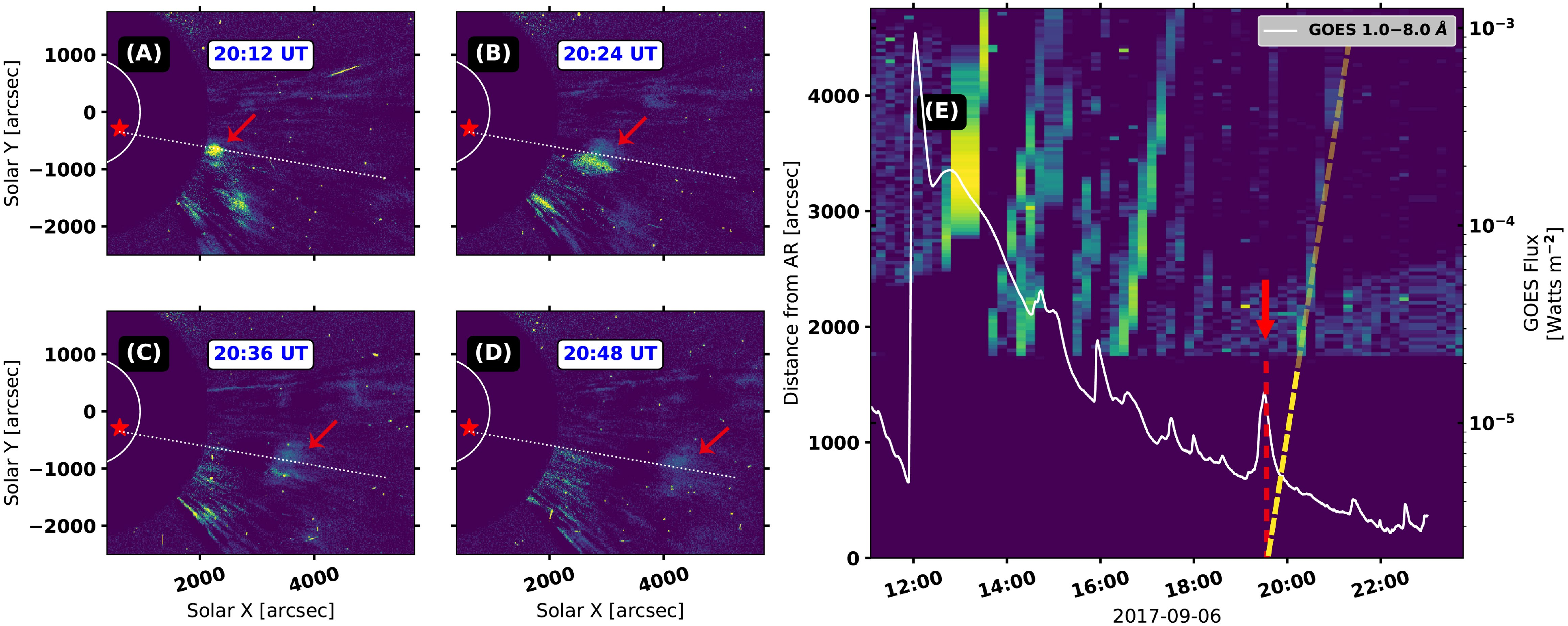}
\caption{\label{fig:lasco} (A)--(D) Running difference image of LASCO C2 coronagraph at selected times showing a narrow eruptive feature in the upper corona associated with the M1.4 flare event (red arrows). (E) Time--distance plot of LASCO C2 running difference image series obtained from the slice shown as a dotted line in (A)--(D). GOES light curve of the day is also shown for reference. The vertical red dashed line indicates the time when the event occurs. The yellow dashed line traces the eruption and extrapolates it back to time of the M1.4 flare that peaks at 19:29 UT. }
\end{figure*}

\subsection{Partial Eruption of a Pre-existing Filament}
At 19:00 UT, when the GST observation of this event starts, a dark filament can be clearly identified in the H$\alpha$ line center images with many strands that appear twisted (Figure~\ref{fig:filament}(A)). The filament can also be distinguished in SDO/AIA 304 \AA\ images albeit with a much lower angular resolution (Figure~\ref{fig:filament}(E)). The filament aligns with the magnetic polarity inversion line (PIL) as evidenced in the SDO/HMI \edit1{radial field} magnetogram (Figure~\ref{fig:story}(D)). \edit1{The magnetogram shows the radial field component $B_r$ derived from the full vector magnetic field measurements (which mitigates the projection effect; see \citealt{2013arXiv1309.2392S})}
The filament corresponds to highly sheared magnetic field lines near the PIL (colored curves in Figure~\ref{fig:story}(D)) derived from the NLFFF results based on the pre-event SDO/HMI vector magnetogram.

The flare enters its impulsive phase at around 19:22 UT. The southern tips of both ribbons brighten first (\edit1{Figure~\ref{fig:filament}(J)--(L)}). The post-reconnection flare arcade that connects the southern tips of the ribbons is clearly seen in SDO/AIA channels sensitive to hot flaring plasma, as shown in Figure~\ref{fig:story}(F). During this period, an upward motion of the filament is also observed \edit1{(see Figure~\ref{fig:filament} and the animation accompanying Figure~\ref{video:sdo_gst})}. Figure~\ref{fig:stack}(C) shows the time-distance diagram derived from the H$\alpha$ time-series images made at a slice that is nearly perpendicular to the filament axis (thick curve in Figure~\ref{fig:stack}(B)). Synchronous with the onset of the impulsive phase of the flare, the dark filament starts to rise with a projected speed of $\sim$23~km s$^{-1}$ (Figure~\ref{fig:story}(A)). Meanwhile, the 6--13 KeV RHESSI source appears near the top of the bright EUV flare arcade.
X-ray spectral analysis suggests that the source is associated with thermal bremsstrahlung emission from $\sim$28 MK plasma (not shown here).

The multi-wavelength observations during the impulsive phase are generally consistent with the standard scenario of eruptive flares as illustrated in Figures~\ref{fig:story}(B) and (C). The rising filament, likely the lower portion of a twisted MFR, stretches the overlying field lines, leading to magnetic reconnection below the filament/MFR. The energy release associated with the reconnection results in a bright EUV flare arcade and a looptop X-ray source as shown in Figures~\ref{fig:story}(E) and (F).

During the impulsive phase of the flare, at $\sim$19:26:20 UT, the rising H$\alpha$ filament appears to go through a fast eruption and quickly disappears from the field of view of BBSO/GST. Immediately following the eruption, the filament material starts to drain toward either end. Owing to the relatively small scale of the eruption, it is rather difficult to trace the erupted filament material into higher altitudes. However, we have identified a small and narrow CME in the SOHO/LASCO C2 running difference images, as shown in Figures~\ref{fig:lasco}(A)--(D). Both the initiation time and location of the eruption, after extrapolating the white light CME in the time-distance diagram in the upper corona ($>\!2.5 R_{\odot}$; shown in Figure~\ref{fig:lasco}(E)) back to the solar surface, are consistent with those of the M1.4 flare event. Considering that the filament feature is still distinguishable after the event (but with an altered appearance), we conclude that only a fraction of the filament material has erupted. Hence, after \citet{Gibson2006a, Gibson2006b}, we refer to this event as a ``partial eruption.''

 \section{Microwave Observations}\label{sec:MW}
 
\subsection{Microwave Counterpart of the Erupting Filament}\label{sec:mw_source}
\begin{figure*}
\centering
\includegraphics[width=0.9\textwidth]{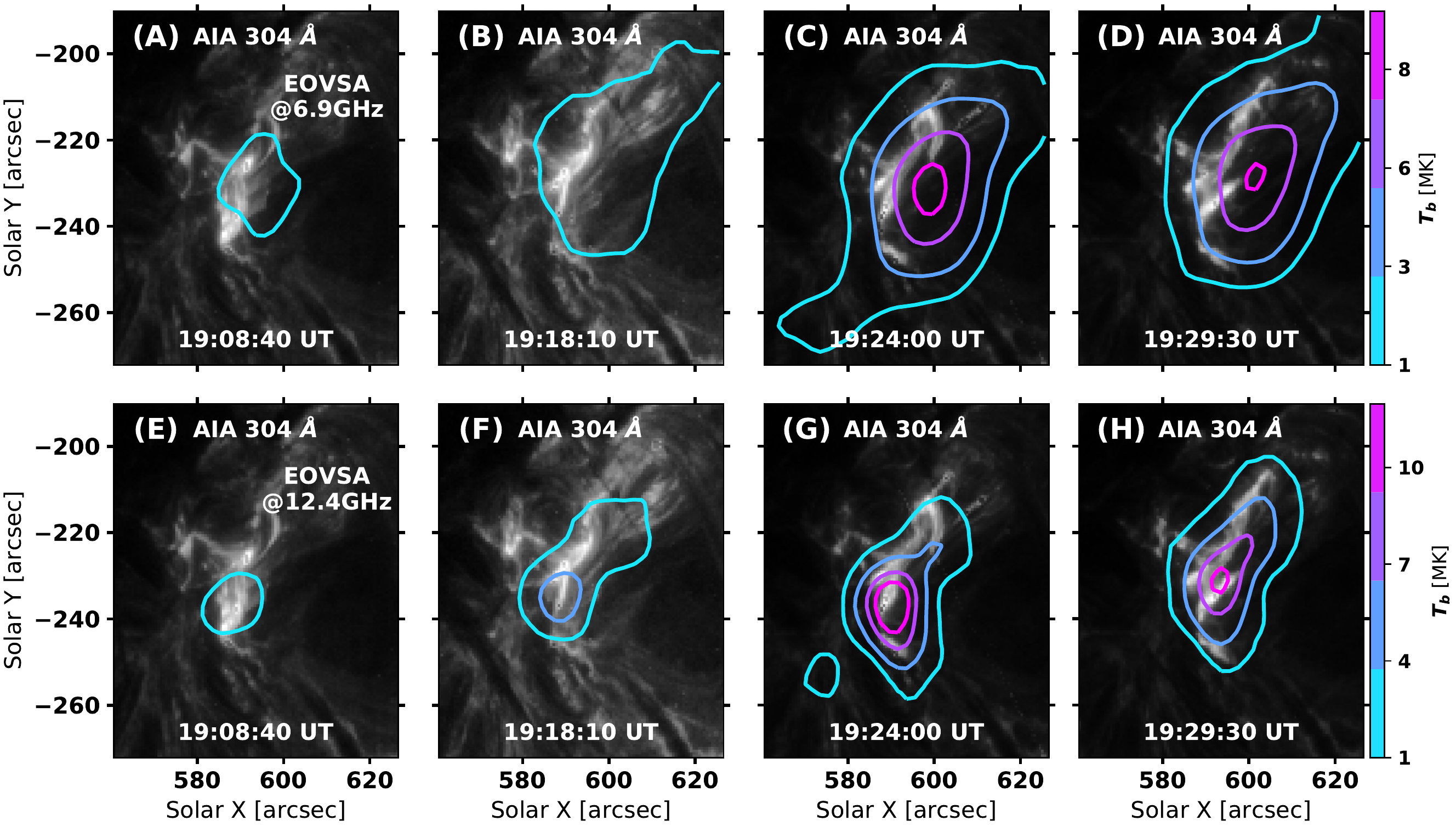}
\caption{\label{fig:mw} Morphology and evolution of EOVSA microwave sources at 6.9 GHz (A)--(D) and 12.4 GHz (E)--(H). Note the contour levels correspond to absolute brightness temperature values shown in the color bars on the right. The background images are from SDO/AIA 304~\AA\ images at the same selected times as those in Figure~\ref{fig:filament}. }
\end{figure*}

In this study, we combine all channels of each of the 30 spectral windows centered at 3.4 to 17.9 GHz (SPW 1--SPW 30) to produce microwave images at 30 equally spaced frequencies. 
The microwave images at all frequencies feature a source near the filament seen in (E)UV and H$\alpha$ images. In Figures \ref{fig:mw}(A)--(H)), we choose two representative frequencies (6.9 GHz and 12.4 GHz) to demonstrate the morphology of the microwave sources. Prior to the flare event, the microwave sources are mainly concentrated in the AR with strong magnetic field (Figures \ref{fig:mw}(A) and (E)), indicative of thermal emission associated with the AR (which will be further discussed in the next sub-section). During the event, the microwave source at both frequencies start to display an elongated shape stretching along the direction of the filament. To display the microwave source morphology at different frequencies more clearly, we perform pre-flare background subtraction on all the microwave images and discuss the resulting images in the subsequent analysis. The kernel of these pre-flare background-subtracted microwave sources, defined as 90\% of the maximum brightness of each image, is closely aligned with the filament (Figures~\ref{fig:story}(C)). Intriguingly, the microwave emission kernels at the different frequencies form a coherent structure that is distributed along the filament, with its high-frequency end located closer to the southern leg of the filament (Figure~\ref{fig:story}(C)).

Moreover, a detailed look at the temporal variation of the source location reveals that the microwave source and the filament also move synchronously during its slow rise phase. 
In Figure~\ref{fig:stack}(C), we show the evolving location of the microwave source kernel (at three selected frequencies) along the same slice used for generating the H$\alpha$ time--distance stack plot. It is clear that the microwave source rises synchronously with the H$\alpha$ filament at a similar speed, $\sim$23 km s$^{-1}$ in projection, albeit with a slight, 5--6$''$ offset toward the direction of the rise motion. Such a close spatial association and synchronized motion between the multi-frequency microwave source kernels and the H$\alpha$/EUV filament strongly suggests that the microwave source is a counterpart of the erupting filament. 
The relation between the microwave emission and the filament will be further discussed and interpreted in Section~\ref{sec:discussion}.

\begin{figure*}
\centering
\includegraphics[width=1.0\textwidth]{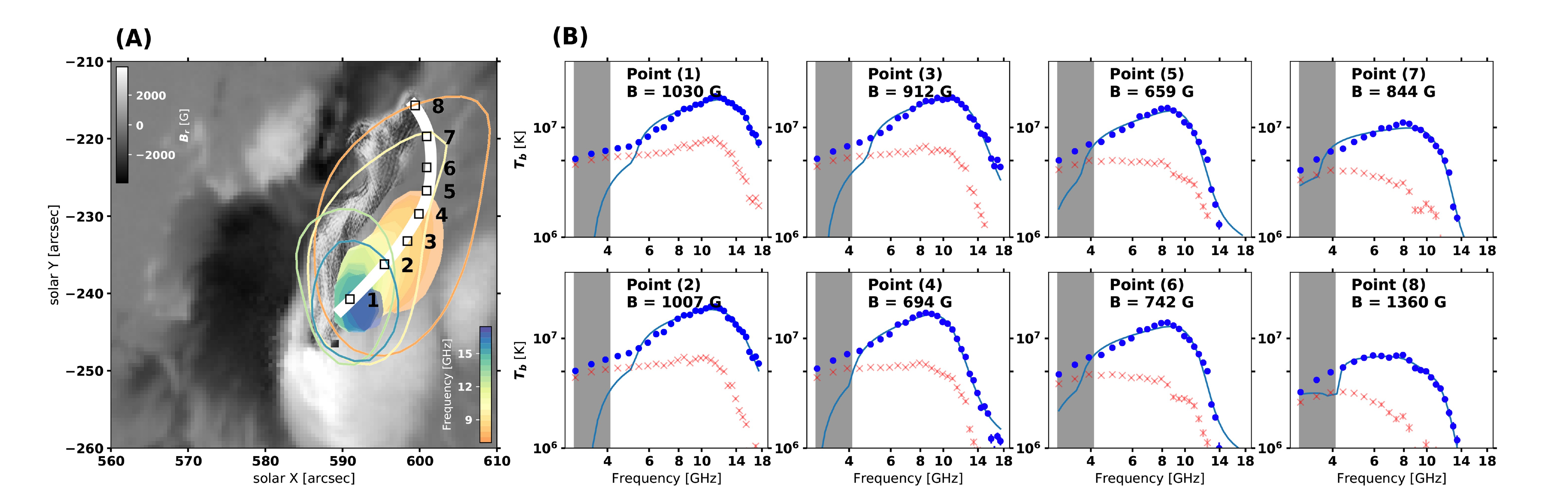}
\caption{\label{fig:spectra} Spatially resolved microwave spectra along the microwave counterpart of the erupting filament. (A) Multi-frequency EOVSA microwave images (pre-flare-background subtracted) at 19:24:00 UT. Open and filled contours are 50\% and 90\% of the maximum, respectively. Background is the pre-flare photospheric magnetogram from HMI, showing the radial component $B_r$. The filament shown is extracted from the BBSO/GST H$\alpha$ image. (B) Spatially resolved microwave brightness temperature spectra derived from selected locations along the extension of the microwave source (black squares). The thick white line indicates the selected magnetic flux tube derived from NLFFF extrapolations, which is used to compare with the results obtained by microwave spectral analysis. The spatially resolved spectra and the best fit model are shown as filled blue circles and solid blue curves, respectively. For comparison, the pre-flare microwave spectra at the same locations are shown as the red crosses.}
\end{figure*}

\subsection{Microwave Spectral Analysis}

In Section~\ref{sec:mw_source}, we have suggested that the microwave sources are the counterpart of the erupting filament. 
To investigate the physical parameters of the microwave source region, we derive spatially resolved microwave brightness temperature spectra obtained at different spatial locations along the elongation direction of the microwave source (black boxes in Figure~\ref{fig:spectra}(A)). Figure~\ref{fig:spectra}(B) shows the brightness temperature spectra obtained from eight selected locations during the first microwave peak at 19:24 UT (solid blue circles). The pre-flare spectra, obtained from the same locations but at 19:03:10 UT, are shown as the red crosses. 
We note that the spectra at frequencies below 4.5 GHz show little enhancement during the flare. They have a brightness temperature of $\sim$7 MK and display a nearly flat spectral shape. The corresponding microwave images are also very extended, encompassing almost the entire active region. We suspect that this spectral regime has a significant contribution from the background thermal emission from the active region, possibly enhanced by the previous X9.3 event. Therefore, in our spectral analysis, we have excluded the data points at $<$4.5 GHz (shaded gray in Figure~\ref{fig:spectra}B). At $>$4.5 GHz, the microwave spectra show a prominent increase during the flare, suggestive of their intimate relationship to the flare energy release. The spectral shape has a positive slope below a peak frequency of $\sim$8--10 GHz and a negative slope above the peak, characteristic of the nonthermal gyrosynchrotron radiation \citep{1985ARA&A..23..169D, Gary2004}.

\begin{figure*}
\centering
\includegraphics[width=0.8\textwidth]{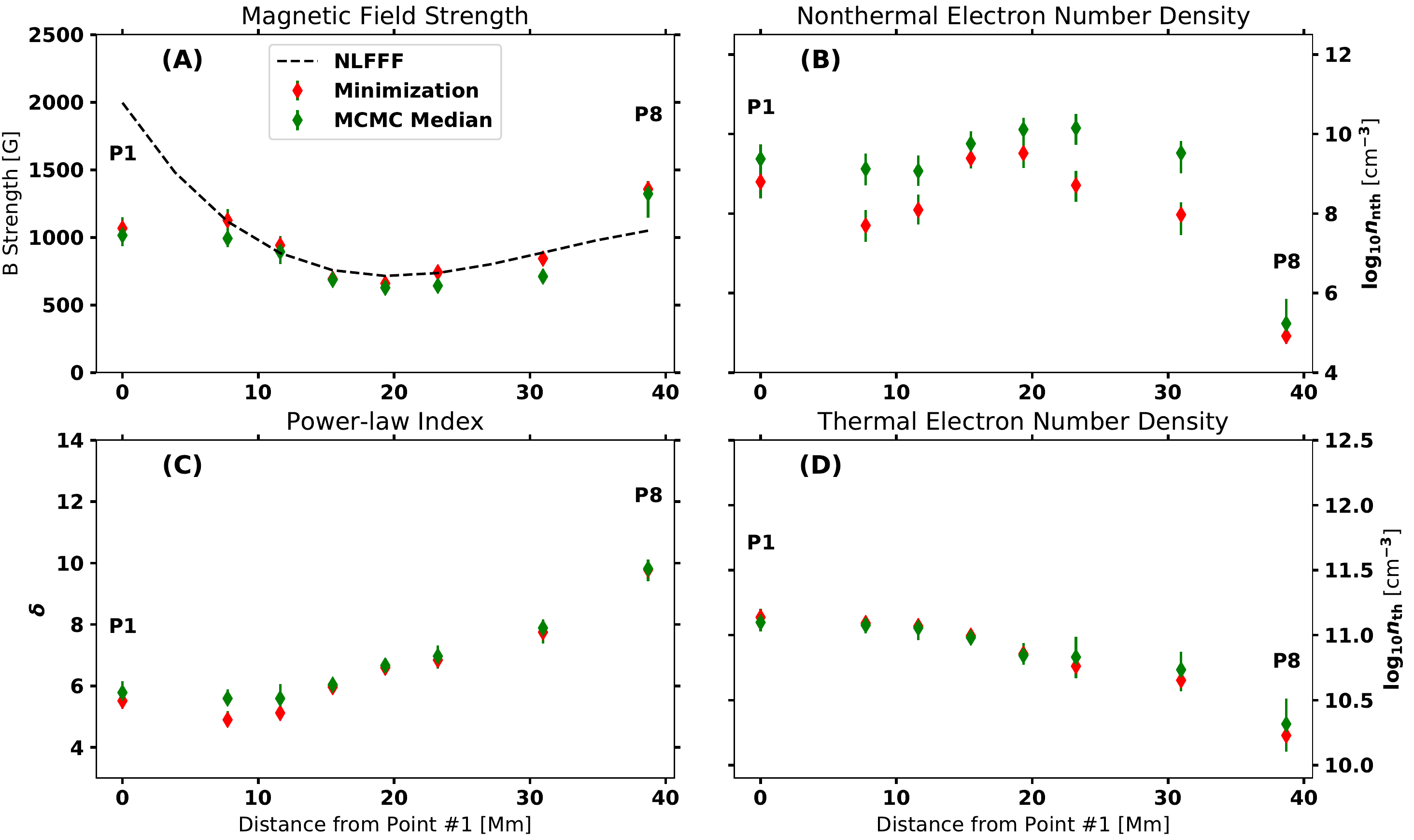}
\caption{\label{fig:params} Spatial variation of key fit parameters along the microwave counterpart of the erupting filament. 
Panels (A)--(D) show, respectively, the magnetic field strength $B$, nonthermal electron density $n_{\rm nth}$, power-law index of the electron energy distribution $\delta$, and thermal plasma density $n_{\rm th}$. Red and green symbols denote those constrained from the $\chi$-square minimization and MCMC, respectively. Also shown in (A) is the magnetic field strength derived from a magnetic flux tube that passes the eight selected locations in the NLFFF results (dashed black curve).}
\end{figure*}

We use the fast gyrosynchrotron code of \cite{Fleishman2010} to calculate the nonthermal microwave emission by assuming a homogenous source along the line of sight (LOS) with a power-law electron energy distribution. After \citet{Fleishman2020}, we adopt a downhill simplex method (implemented in SciPy's \citep{2020SciPy-NMeth} \texttt{minimize} package as the ``Nelder-Mead'' algorithm) to minimize the $\chi$-square differences between the observed and modeled gyrosynchrotron spectra. For the $\chi$-squared minimization based spectral fit, four free parameters are used, which include the magnetic field strength $B$, the total number density of nonthermal electron $n_{\rm nth}$, the power-law index of the electron energy distribution $\delta$, and the thermal electron density $n_{\rm th}$. The column depth is fixed to 10$''$, a value assumed based on the source size in the plane of the sky. The energy range of the power-law distribution is fixed to 10 keV--10 MeV, and the temperature of the thermal plasma is fixed to 7~MK. Following \citet{Chen2020a} and \citet{2021ApJ...908L..55C}, we also adopt the Markov chain Monte Carlo (MCMC) method to evaluate the reliability and uncertainties of the fit parameters. The best-fit spectra are shown in Figure~\ref{fig:spectra}(B) as the solid blue curves. The corresponding best-fit parameters of the eight selected locations are shown in Figure~\ref{fig:params}. For completeness, we also show the median values and the associated 1-$\sigma$ range of the MCMC posterior distributions as the green symbols.   

The spectral fit results return a spatially varying magnetic field strength $B$ along the microwave counterpart of the filament, shown in Figure~\ref{fig:params}(A) as red and green symbols (from the minimization and MCMC median, respectively): It decreases from $\sim$1000~G at the southern end of the source to $\sim$600~G near the center, and then increases to $>$1000~G at the northern end. To compare this microwave-constrained magnetic field distribution with the NLFFF results, we extract the magnetic field strength values from an NLFFF-extrapolated magnetic flux tube that passes the selected fit locations in projection, shown in Figure~\ref{fig:params}(A) as the black dashed curve. The two results achieve a qualitative agreement with each other, despite some deviations at either end. The latter may be attributed to the projection effect and/or temporal evolution of the coronal magnetic field from the pre-flare phase when the NLFFF results are derived. 

The spectral index of the electron distribution $\delta$ and nonthermal electron density $n_{\rm nth}$ also vary along the microwave source, with a greater $n_{\rm nth}$ and a harder $\delta$ near the southern leg of the MFR. It is consistent with the observations presented in Section~\ref{sec:multi}, where the bright EUV flare arcade appears near the southern end of the PIL, indicating that the energy release and the associated electron acceleration may be more profound there.

\section{Summary and Discussions}\label{sec:discussion}
In the previous sections, we have presented microwave, H$\alpha$, and (E)UV observations of a filament that undergoes a partial eruption during an M1.4-class solar flare. In particular, the multi-frequency microwave counterpart of the filament closely follows the morphology and dynamics of the rising filament as seen in H$\alpha$ and EUV. By fitting the spatially resolved microwave spectra using a gyrosynchrotron radiation model, we derive the magnetic field strength along the filament, which ranges from 600--1400 G from its apex to the legs. The microwave-constrained magnetic field yields a reasonable agreement with those derived from the NLFFF extrapolation. These results strongly suggest that the observed microwave, H$\alpha$, and EUV features are all closely associated with the same coherent magnetic structure that hosts the filament, presumably a twisted MFR that undergoes a partial eruption. 

\begin{figure*}
\centering
\includegraphics[width=0.8\textwidth]{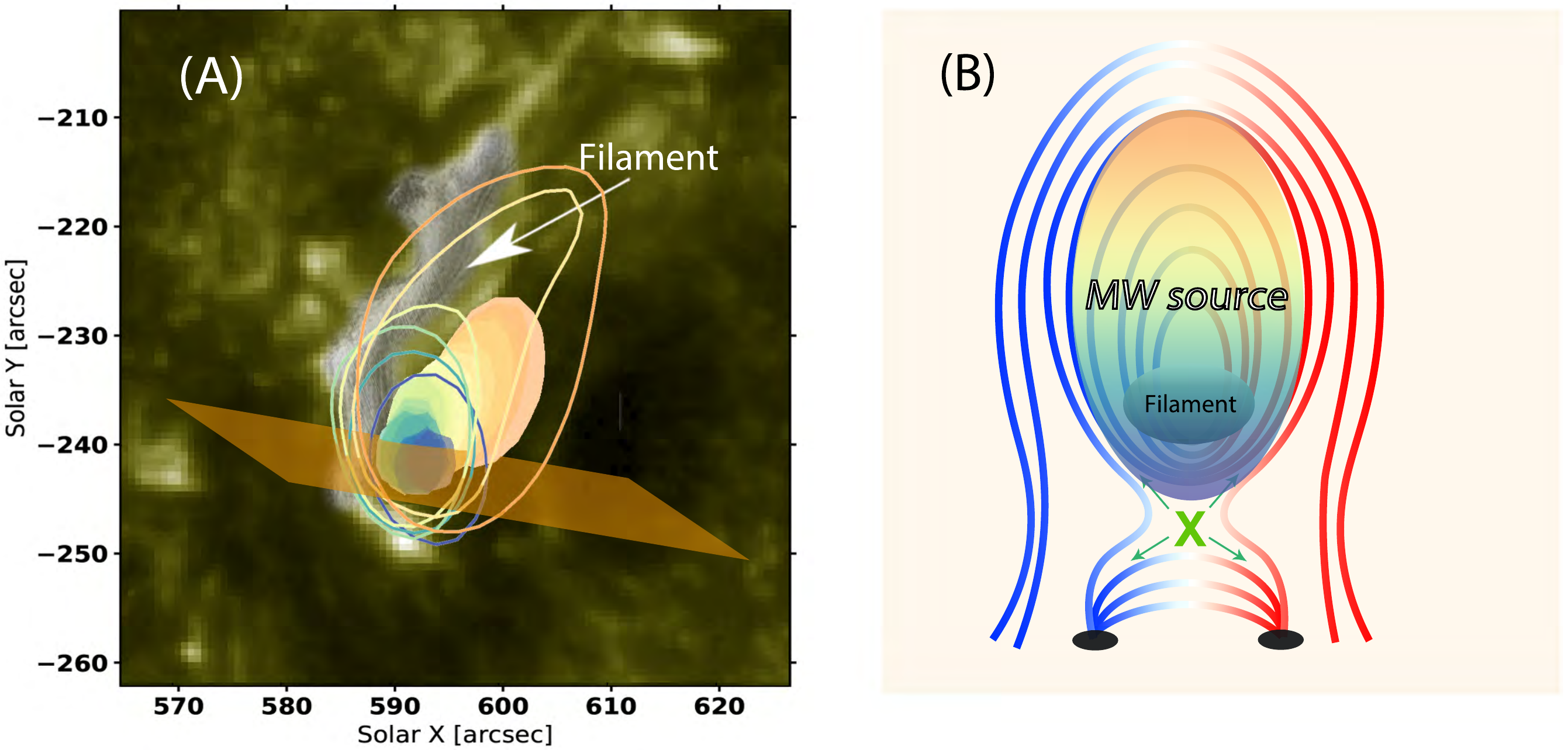}
\caption{\label{fig:filament_cartoon} Schematic cartoon that shows the relationship between the filament and the microwave source. (A) Multi-frequency EOVSA microwave images (pre-flare-background subtracted) at 19:24:00 UT. Open and filled contours are 50\% and 90\% of the maximum, respectively. Background is the SDO/AIA 1600~\AA\ image at the moment and the filament shown is extracted from the BBSO/GST H$\alpha$ image. (B) Schematic cartoon of the cross-section of the flux rope, at a location indicated by the orange surface in (A).}
\end{figure*}

It is particularly intriguing that although the multi-frequency microwave source encompasses the filament, the centroid of the microwave source is located consistently above the rising filament as seen in H$\alpha$/EUV (Figures~\ref{fig:stack}(C)).
Figure~\ref{fig:filament_cartoon}(A) shows a representative frame at 19:24:00 UT with EOVSA microwave sources overlaid on the composite AIA 1600 \AA\ (green background) and BBSO/GST H$\alpha$ (grayscale) image. Similar to Figure~\ref{fig:spectra}(A), open and filled color contours are 50\% and 90\% of the microwave source, showing, respectively, the spatial extension and the central kernel of the source at each frequency.
We note the viewing geometry of this event is nearly top-down but slightly tilted toward the west. In Figure~\ref{fig:filament_cartoon}(B), we show a schematic of the cross-section of the filament--MFR system (indicated by the orange plane in (A)). The relative location and cross-section of the microwave source and the filament are illustrated by the green-yellow and gray ellipse, respectively. The position difference between the microwave source and the H$\alpha$ filament can be understood within the standard scenario of the ``three-part'' structure of the filament--MFR system in conjunction with the reconnection-driven flare energy release associated with the (partial) filament eruption \citep[e.g.,][and references therein]{Gibson2006b, Dove2011, 2014ApJ...794..149C}: The cool, dense filament observed in H$\alpha$ and EUV 304 \AA\ can be explained as chromospheric-temperature material supported near the concave-upward bottoms of the field lines of an MFR. Meanwhile, accelerated electrons due to magnetic reconnection induced by the partial filament eruption can enter the extended MFR/CME cavity following the newly reconnected field lines \citep[see, e.g.,][]{2013ApJ...779L..29G,Chen2020a}, producing the extended nonthermal microwave source \edit1{above the rising filament. We note that similar phenomena have also been observed in the upper corona using radio data obtained at longer wavelengths: there have been reports of moving type IV radio bursts or ``radio CMEs'' located ahead of the erupting filament \citep{Vrsnak2003a} or accompanying the extended CME cavity \citep{2001ApJ...558L..65B,2007ApJ...660..874M,2017A&A...608A.137C,2020ApJ...893...28M,2021ApJ...906..132C}.} 

Because the microwave intensity depends strongly on both the nonthermal electron distribution and the magnetic field strength, the slight offset of the kernel of the microwave emission as a function of frequency relative to the filament contains important information of the magnetic structure and nonthermal electron distribution. \edit1{Also, the nonthermal-to-thermal electron fraction in the microwave source (which can be up to 10\%; c.f., Figures~\ref{fig:params}(B) and (D)) may provide diagnostics for the acceleration processes.} However, \edit1{since the nonthermal electrons responsible for the observed microwave sources are probably accelerated elsewhere}, they cannot be understood straightforwardly without detailed modeling of electron acceleration and transport in the eruption-induced magnetic reconnection geometry. An in-depth interpretation for such observed phenomenon is a topic for future studies that incorporates macroscopic plasma and particle modeling.

\begin{figure}[!ht]
\centering
\includegraphics[width=0.45\textwidth]{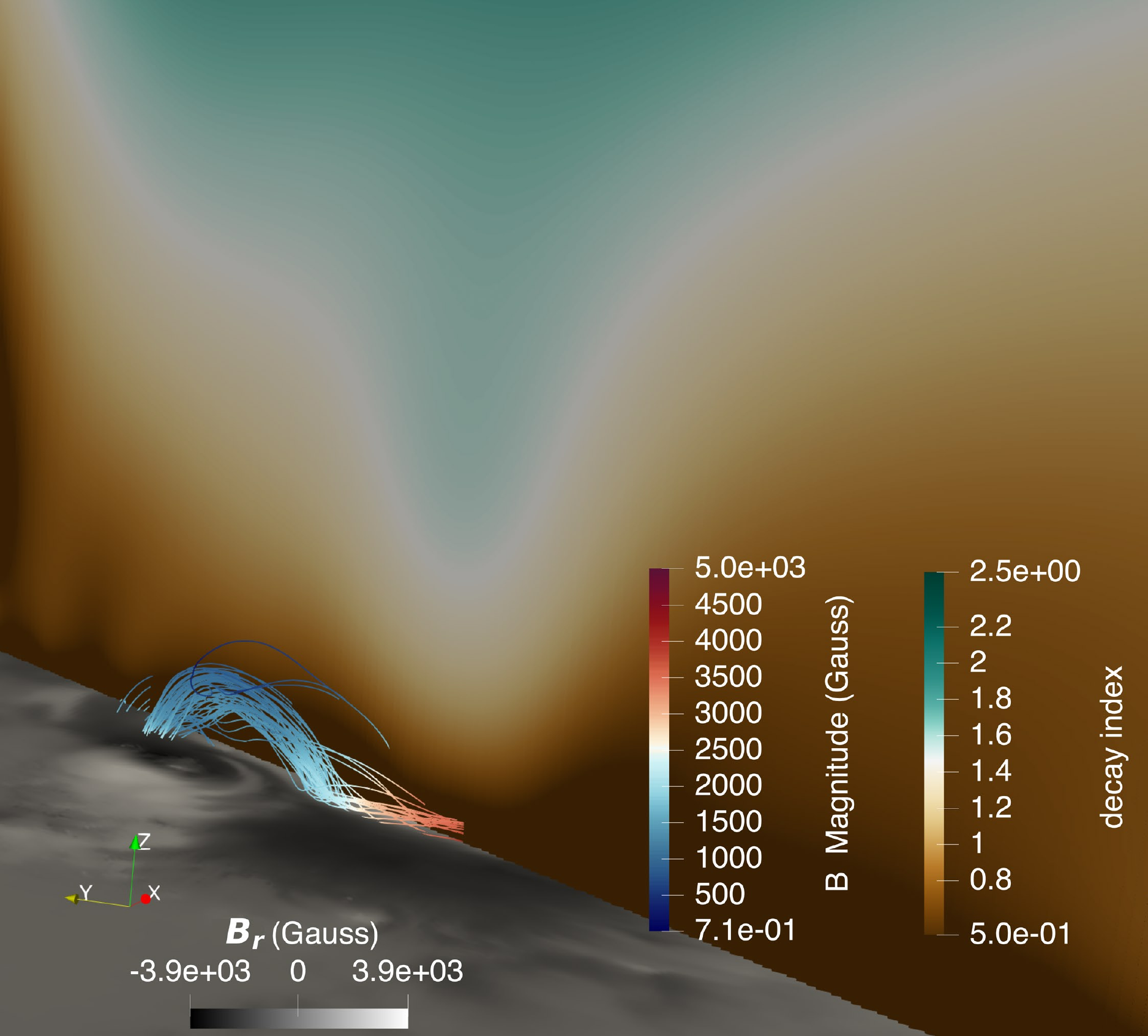}
\caption{\label{fig:decay_index} Decay index distribution above the flux rope. The decay index is calculated from the potential extrapolated magnetic field while the flux rope is from the NLFFF extrapolation.}
\end{figure}

Finally, we offer a possible interpretation for the partial eruption of the filament–MFR system under the framework of the torus instability. Following earlier works \citep[e.g.,][]{Bateman1978, Kliem2006}, we calculate the potential coronal magnetic field above the AR, and compute the decay index using the transverse component of the pre-flare potential field. 
Figure~\ref{fig:decay_index} shows the decay index map in a vertical plane that aligns with the main filament axis (nearly perpendicular to the orange plane in Figure~\ref{fig:filament_cartoon}(A)). The regions colored in brown have a decay index of $<$1.5, which are stable against the torus instability. In contrast, the regions colored in green have a decay index of $>$1.5, where the MFR is unstable to the torus instability and hence is more likely to erupt. It can be seen from the decay index map that the MFR is largely stable against the torus instability. However, once the filament is activated and driven upward by, e.g., the Lorentz force from pre-flare reconnection events \citep{jiang2021}, the southern portion of the filament (to the right in the diagram) can quickly ascend into a region in the torus-unstable regime. However, the northern portion of the filament has much more difficulty erupting owing to the more extended torus-stable region above the MFR. Such a north-south asymmetry may explain the observed partial eruption of the filament and the concentration of the nonthermal microwave source near its southern end.

To briefly summarize, by combining multi-wavelength observations from BBSO/GST, SDO, RHESSI and, in particular, microwave imaging spectroscopy observations from EOVSA, we provide the first measurement of the spatially resolved magnetic field along an erupting filament in a flare-productive AR. The microwave-constrained results are qualitatively consistent with those derived from the NLFFF extrapolation. Our study demonstrates the unique role of microwave imaging spectroscopy observations in measuring the dynamic magnetic field and accelerated electrons on the active Sun. However, the limited angular resolution, dynamic range, and image fidelity of EOVSA observations of this event do not allow us to derive a detailed map of the magnetic field distribution above the filament. Such a magnetic map would provide the most direct constraints for understanding the eruption conditions including the decay index. These measurements should be routinely available with a next generation solar radio telescope with improved spatial resolution, such as the Frequency Agile Solar Radiotelescope \citep{Bastian2019}.  

\acknowledgements  

This work is supported by NSF under grants AGS-1654382, AGS-1954737, AGS-1821294, and NASA under grants 80NSSC19K0068, 80NSSC20K0627, 80NSSC20K1282, 80NSSC20K1318, 80NSSC21K1671, and 80NSSC21K0003 to NJIT. EOVSA operation is supported by NSF grants AST-1910354 and AGS-2130832. BBSO operation is supported by NJIT and NSF AGS-1821294. GST operation is partly supported by the Korea Astronomy and Space Science Institute, the Seoul National University, and the Key Laboratory of Solar Activities of Chinese Academy of Sciences (CAS) and the Operation, Maintenance and Upgrading Fund of CAS for Astronomical Telescopes and Facility Instruments. The authors acknowledge Gelu Nita and Gregory Fleishman for help with the GX simulator package. 

\software{SciPy \citep{2020SciPy-NMeth},
          AstroPy \citep{Robitaille2013}, 
          CASA \citep{2007ASPC..376..127M},
          Lmfit \citep{Newville2016},
          MGN \citep{Morgan2014},
          NumPy \citep{harris2020array},
          SunPy \citep{Community2015},
          GX simulator \citep{Nita2015}}

\vspace{1.5in}

\appendix
An interactive animation similar to Figure~\ref{fig:filament} in the main text is included below. The animation shows the evolution of the H$\alpha$, 304~\AA\, and 1600~\AA\ images during the SOL2017-09-06T19:29:30 M1.4 flare. Also shown in the animation are the microwave dynamic spectrum and X-ray light curves.

\begin{figure*}[!ht]
\begin{interactive}{animation}{sdo_gst_animation.mp4}
\plotone{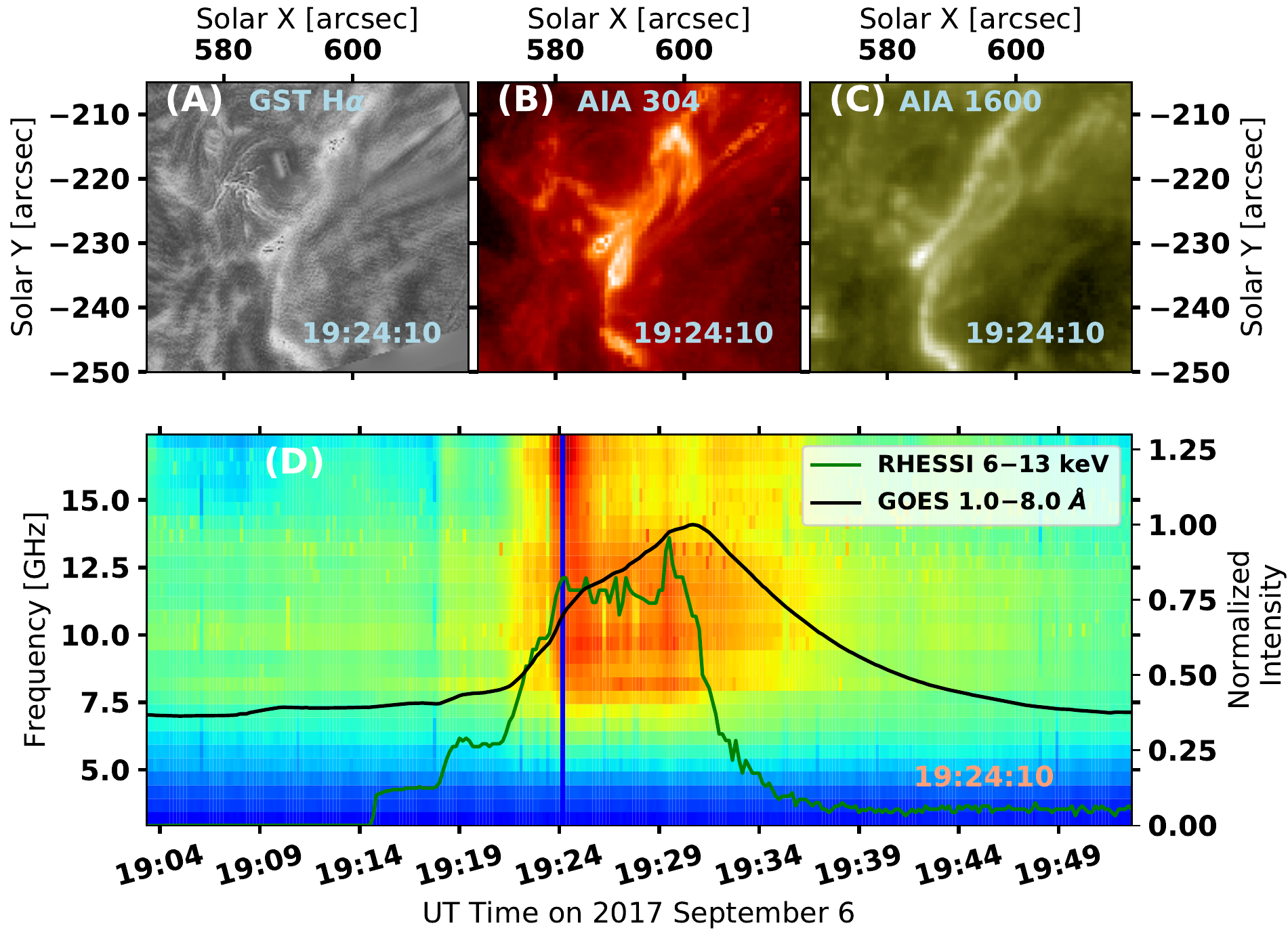}
\end{interactive}
\caption{(A)--(C) Animation of BBSO/GST H$\alpha$ line center images, SDO/AIA 304~\AA\ images, and SDO/AIA 1600~\AA\ images, respectively. (D) Background-subtracted EOVSA microwave dynamic spectrum from 19:03 UT to 19:52 UT. The black and green curves are for RHESSI 6--13 keV X-ray and GOES 1--8~\AA\ light curves, respectively. The vertical blue line indicates the closest time of the animation frame. The animation covers 49 minutes of observing beginning at 19:03 UT on 2017 September 6. The duration of the animation is 15 s.}
\label{video:sdo_gst}
\end{figure*}

\bibliography{Wei2021}{}
\bibliographystyle{aasjournal}


\end{document}